\documentclass[a4paper]{article}

\usepackage{INTERSPEECH2016}
\usepackage{graphicx,amssymb,amsmath,bm,textcomp,subfigure,multirow,epstopdf,graphicx,tabularx}

\ninept

\title{Domain adaptation based Speaker Recognition on Short Utterances}

\makeatletter
\def\name#1{\gdef\@name{#1\\}}
\makeatother 
\name{{\em Ahilan Kanagasundaram$^{*+}$, David Dean$^{*}$, Sridha Sridharan$^{*}$ and Clinton Fookes$^{*}$}}
\address{Speech and Audio Research Laboratory$^{*}$  \\
	Queensland University of Technology, Brisbane, Australia$^{*}$ \\
	{\small \tt \{a.kanagasundaram, d.dean, s.sridharan, c.fookes\}@qut.edu.au}$^{*}$ \\
	Electrical \& Electronic Engineering, Faculty of Engineering$^{+}$ \\
	University of Jaffna, Jaffna, Sri Lanka$^{+}$\\
	{\small \tt ahilan@eng.jfn.ac.lk}$^{+}$
}

\begin{document}

\maketitle
\begin{abstract}
This paper explores how the in- and out-domain probabilistic linear discriminant analysis~(PLDA) speaker verification behave when enrolment and verification lengths are reduced. Experiment studies have found that when full-length utterance is used for evaluation, in-domain PLDA approach shows more than 28\% improvement in EER and DCF values over out-domain PLDA approach and when short utterances are used for evaluation, the performance gain of in-domain speaker verification reduces at an increasing rate. Novel modified inter dataset variability~(IDV) compensation is used to compensate the mismatch between in- and out-domain data and IDV-compensated out-domain PLDA shows respectively 26\% and 14\% improvement over out-domain PLDA speaker verification when SWB and NIST data are respectively used for S normalization. When the evaluation utterance length is reduced, the performance gain by IDV also reduces as short utterance evaluation data i-vectors have more variations due to phonetic variations when compared to the dataset mismatch between in- and out-domain data.
\end{abstract}

\section{Introduction}
In a typical speaker verification system, the significant amount of speech is required for speaker model enrolment and verification, especially in the presence of large intersession variability. Techniques based on factor analysis, such as joint factor analysis~(JFA)~\cite{Kenny2005,Vogt2008b}, i-vectors~\cite{Dehak2010} and probabilistic linear discriminant analysis~(PLDA)~\cite{Kenny2010}, have demonstrated outstanding behaviour in National Institute of Standards and Technology~(NIST) conditions~\cite{NIST2008,NIST2010}. Unfortunately, the performance of many of these approaches degrades rapidly as the available amount of enrolment and/or  verification speech decreases~\cite{Vogt2008a,Kanagasundaram2011,Kanagasundaram2012b}, limiting the utility of speaker verification in real world applications, such as access control or forensics.

Recent studies have found that when speaker verification is developed on data from the Switchboard database and evaluated using data from NIST evaluations, the dataset mismatch significantly affects the speaker verification performance~\cite{Garcia-Romero2014,Aronowitz2014,Glembek}. Several techniques have been proposed to address this issue and achieve state-of-the-art speaker verification performance when speaker verification developed in one domain data and evaluated in different domain data.

The main aim of this paper is to investigate the effect of only having short utterances available for evaluation for the out- and in-domain PLDA speaker verification. Previous studies have shown that when PLDA speaker verification is developed using out-domain data and evaluated using in-domain long utterances, it fails to achieve good performance due to mismatch between out- and in-domain data. In this paper, it is analysed whether the above outcome is true with short utterance evaluation data as well. Subsequently, novel modified inter dataset variability~(IDV) compensation approach is studied with short utterance evaluation data in order to find whether those approaches improve the performance of out-domain short utterance PLDA speaker verification system. We also analyse the speaker verification performance with regards to the duration of utterances used for both speaker evaluation (enrolment and verification) and score normalization.

This paper is structured as follows: Section~\ref{sec:i-vector feaure extraction} details the i-vector feature extraction techniques. Section~\ref{sec:IDV compensation} details the inter dataset variability compensation approach. Section~\ref{sec:len-norm GPLDA} explains the Gaussian PLDA~(GPLDA) based speaker verification system. The experimental protocol and corresponding results are given in Section~\ref{sec:method} and Section~\ref{sec:results and discussions}. Section~\ref{sec:conclusion} concludes the paper.
\section{I-vector feature extraction} \label{sec:i-vector feaure extraction}
I-vectors represent the Gaussian mixture model~(GMM) super-vector by a single total-variability subspace. This single-subspace approach was motivated by the discovery that the channel space of JFA contains information that can be used to distinguish between speakers~\cite{Dehak2009a}. An i-vector speaker and channel dependent GMM super-vector can be represented by,
\begin{eqnarray}
\boldsymbol{\mu} & = & \textbf{m} + \textbf{Tw},
\end{eqnarray}
where $\textbf{m}$ is the same universal background model~(UBM) super-vector used in the JFA approach and $\textbf{T}$ is a low rank total-variability matrix. The total-variability factors~($\textbf{w}$) are the i-vectors, and  are normally distributed with parameters $\emph{N(0,1)}$. Extracting an i-vector from the total-variability subspace is essentially a \textit{maximum a-posteriori adaptation}~(MAP) of $\textbf{w}$ in the subspace defined by $\textbf{T}$. An efficient procedure for the optimization of the total-variability subspace $\textbf{T}$ and subsequent extraction of i-vectors is described Dehak~\emph{et al.}~\cite{Dehak2010,Kenny2008}. In this paper, the pooled total-variability approach is used for i-vector feature extraction where the total-variability subspace~(${R_{w}}^{telmic} = 500$) is trained on SWB dataset.
\section{Modified IDV compensation approach} \label{sec:IDV compensation}
When PLDA speaker verification is trained using out-domain data, it significantly affects the speaker verification performance due to the mismatch between evaluation and development data. Recently, we proposed inter dataset variability~(IDV) compensation approach to compensate the mismatch between out- and in-domain data~\cite{Kanagasundaram2015(Submitted)}. In this paper, we propose modified IDV compensation approach to effectively compensate the mismatch between out- and in-domain data. In previous IDA approach, dataset mismatch variation was calculated using the outer product of the difference between the out-domain i-vectors and average of speaker unlabelled in-domain i-vectors. In this new approach, dataset mismatch between in- and out-domain data is estimated using outer product of difference between the out-domain i-vectors and average of speaker unlabelled in-domain i-vectors, and outer product of difference between in-domain i-vectors and average of speaker unlabelled out-domain i-vectors. The dataset mismatch variation, $\textbf{S}_{IDV}^{'}$, can be calculated as follows,
\begin{eqnarray}
\textbf{S}_{IDV}^{'} =  \frac{1}{N_{1}}\sum_{i=1}^{N_{1}}(\textbf{w}^{OD}_{i} - \bar{\textbf{w}}^{ID})(\textbf{w}^{OD}_{i} - \bar{\textbf{w}}^{ID})^{T} \notag \\
+ \frac{1}{N_{2}}\sum_{j=1}^{N_{2}}(\textbf{w}^{ID}_{j} - \bar{\textbf{w}}^{OD})(\textbf{w}^{ID}_{j} - \bar{\textbf{w}}^{OD})^{T}
\label{eqn:S_IDV new estimation} 
\end{eqnarray}
where $\textbf{w}^{OD}_{i}$ and $\textbf{w}^{ID}_{j}$ are out- and in-domain i-vectors. $\bar{\textbf{w}}^{OD}$ and $\bar{\textbf{w}}^{ID}$ are mean of out- and in-domain i-vectors. $N_{1}$ and $N_{2}$ are number of out- and in-domain i-vectors. SWB data was chosen as out-domain data and 150 speaker with 10 sessions were randomly selected from NIST 2004, 2005 and 2005 telephone data as in-domain data. The decorrelated matrix, $\textbf{D}^{'}$, is calculated using the Cholesky decomposition of $\textbf{D}^{'}{\textbf{D}^{'}}^{T} = \frac{1}{\textbf{S}_{IDV}^{'}}$. Inter dataset variability compensated out-domain i-vectors are extracted as follows,
\begin{eqnarray}
\textbf{w}_{modified-IDV} & = & {\textbf{D}^{'}}^{T}\textbf{w} \label{eqn: IDV projection1}
\end{eqnarray}
Once inter-dataset variability compensated i-vectors, LDA projection is applied to compensate the additional session variation prior to the PLDA modelling and reduce the dimensionality~\cite{Kanagasundaram2013}, which is explained in following in Section~\ref{sec:LDA projection}.
\subsection{LDA approach} \label{sec:LDA projection}
The LDA transformation is estimated based up the standard within- and between-class scatter estimations $S_{b}$ and $S_{w}$, calculated as
\begin{eqnarray}
\textbf{S}_{b} & = & \sum_{s=1}^{S}n_{s}(\bar{\textbf{w}}_{s}-\bar{\textbf{w}})(\bar{\textbf{w}}_{s}-\bar{\textbf{w}})^T, \label{eqn:Sb} \\
\textbf{S}_{w} & = & \sum_{s=1}^{S}\sum_{i=1}^{n_s}(\textbf{w}^{s}_{i}-\bar{\textbf{w}}_{s})(\textbf{w}^{s}_{i}-\bar{\textbf{w}}_{s})^T, \label{eqn:Sw}
\end{eqnarray}
where $S$ is the total number of speakers, $n_{s}$ is number of utterances of speaker $s$. The mean i-vectors, $\bar{\textbf{w}}_{s}$ for each speaker, and $\bar{\textbf{w}}$ is the across all speakers are defined by
\begin{eqnarray}
\bar{\textbf{w}}_{s} = \frac{1}{n_{s}}\sum_{i=1}^{n_s}\textbf{w}^{s}_{i}, \\
\bar{\textbf{w}} = \frac{1}{N}\sum_{s=1}^{S} \sum_{i=1}^{n_s}\textbf{w}^{s}_{i}.
\end{eqnarray}
where $N$ is the total number of sessions. In the first stage, LDA attempts to find a reduced set of axes $\mathbf{A}$ through the eigenvalue decomposition of $\textbf{S}_{b}\textbf{v} = \lambda\textbf{S}_{w}\textbf{v}$.
The IDV-compensated LDA-projected i-vector can be calculated as follows,
\begin{eqnarray}
\hat{\textbf{w}}_{\emph{IDV-LDA}} = {\textbf{A}}^{T}\textbf{w} \label{eqn:LDA projection}
\end{eqnarray}
After LDA-projection, length-normalized GPLDA model parameters are estimated in as described in Section~\ref{sec:len-norm GPLDA}.
\section{Length-normalized GPLDA system} \label{sec:len-norm GPLDA}
\subsection{PLDA modelling}
In this paper, we have chosen length-normalized GPLDA, as it is also a simple and computationally efficient approach~\cite{Garcia-Romero2011}. The length-normalization approach is detailed by Garcia-Romero~\emph{et al.}~\cite{Garcia-Romero2011}, and this approach is applied on development and evaluation data prior to GPLDA modelling. A speaker and channel dependent length-normalized i-vector, $\hat{\textbf{w}}_{r}$ can be defined as,
\begin{eqnarray} \label{PLDA model}
\hat{\textbf{w}}_{r} & = & \bar{\hat{\textbf{w}}} + \textbf{U}_{1}\textbf{x}_{1} + \boldsymbol{\varepsilon}_{r}
\end{eqnarray}
where for given speaker recordings $r = 1,.....R$; $\textbf{U}_{1}$ is the eigenvoice matrix, $\textbf{x}_1$ is the speaker factors and $\boldsymbol{\varepsilon}_{r}$ is the residuals. In the PLDA modeling, the speaker specific part can be represented as $\bar{\textbf{w}} + \textbf{U}_{1}\textbf{x}_{1}$, which represents the between speaker variability. The covariance matrix of the speaker part is $\textbf{U}_{1}{\textbf{U}_{1}}^{T}$. The channel specific part is represented as $\boldsymbol{\varepsilon}_{r}$, which describes the within speaker variability. The covariance matrix of channel part is ${\boldsymbol{\Lambda}}^{-1}$. We assume that precision matrix~($\boldsymbol{\Lambda}$) is full rank. Prior to GPLDA modelling, standard LDA approach is applied to compensate the additional channel variations as well as reduce the computational time~\cite{Kanagasundaram2012a}.
\subsection{GPLDA scoring} \label{sec:PLDA scoring}
Scoring in GPLDA speaker verification systems is conducted using the batch likelihood ratio between a target and test i-vector~\cite{Kenny2010}. Given two i-vectors, $\textbf{w}_{target}$ and $\textbf{w}_{test}$, the batch likelihood ratio can be calculated as follows,
\begin{eqnarray}
\ln\frac{P(\textbf{w}_{target},\textbf{w}_{test}\mid H_{1})}{P(\textbf{w}_{target}\mid H_{0})P(\textbf{w}_{test}\mid H_{0})}
\end{eqnarray}
where $H_{1}$ denotes the hypothesis that the i-vectors represent the same speakers and $H_{0}$ denotes the hypothesis that they do not.
\begin{figure*}
\begin{center}
\begin{tabularx}{\linewidth}{XXX}
\subfigure[EER]{
\includegraphics[width=8cm, height=5cm]{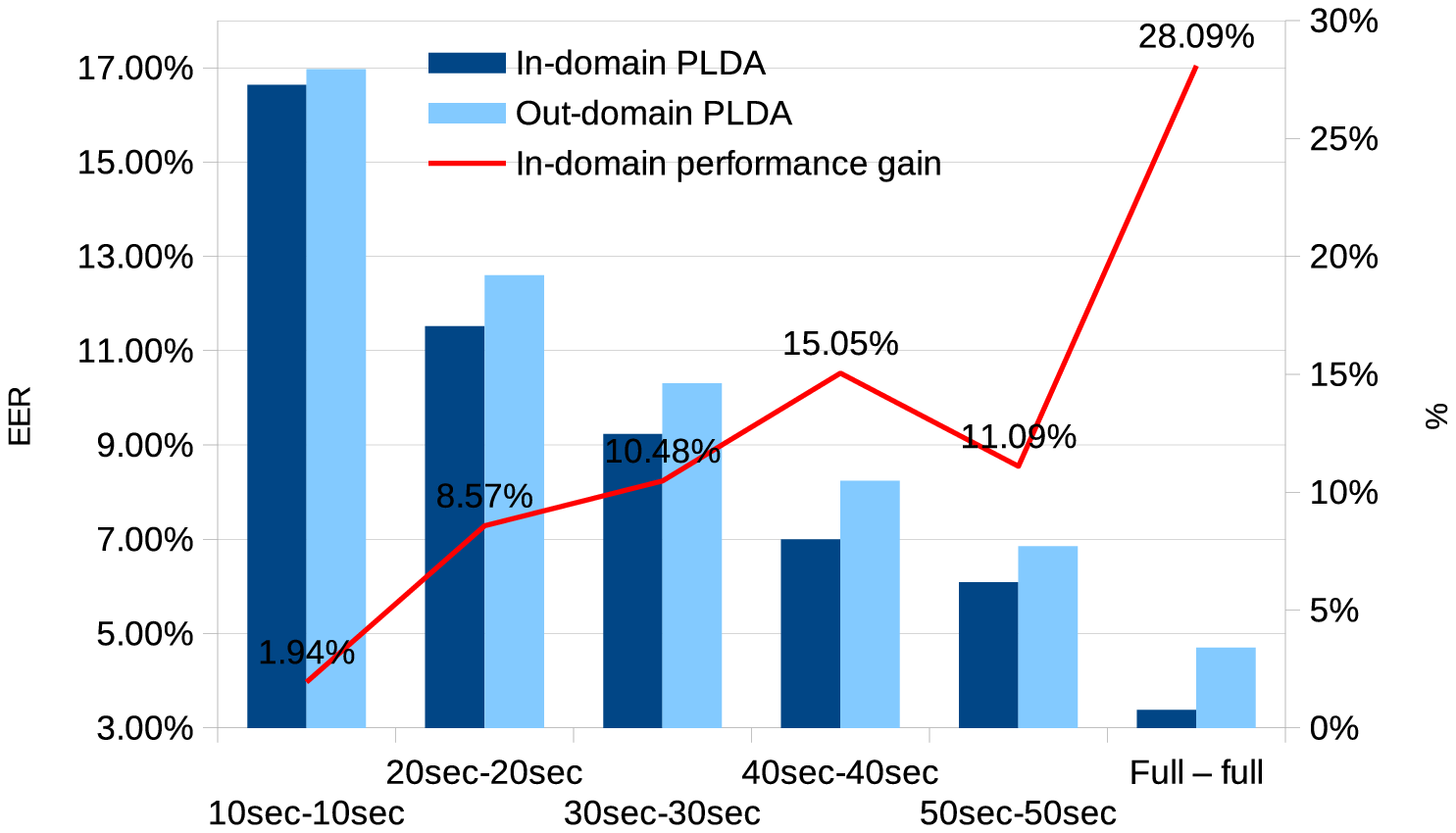}%
\label{fig:EER short eval}%
} &
\subfigure[DCF]{
\includegraphics[width=8cm, height=5cm]{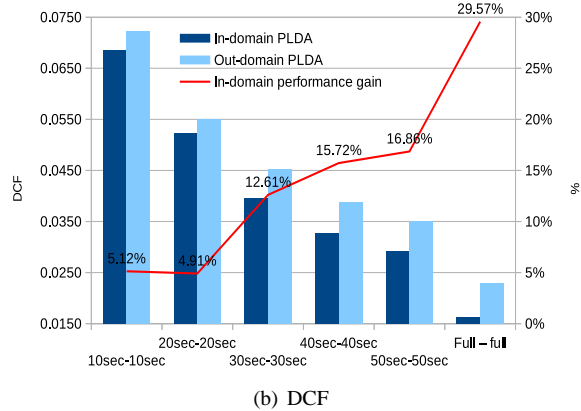}%
\label{fig:DCF short eval}%
}
\end{tabularx}
\end{center}
\caption{\emph{The performance comparison of in- and out-domain PLDA speaker verification on common set of NIST 2008 different short utterance evaluation conditions is shown in bar chart. The performance gain of in-domain PLDA over out-domain PLDA is also shown in line graph} \subref{fig:EER short eval} \emph{EER,} \subref{fig:DCF short eval} \emph{DCF.}}
\label{fig:in and out domain PLDA on short eval}
\end{figure*}
\section{Experimental methodology} \label{sec:method}
The proposed methods were evaluated using the the NIST 2008 SRE corpora. The shortened evaluation utterances were obtained by truncating the NIST 2008 \emph{short2}-\emph{short3} condition to the specified length of active speech for both enrolment and verification. Prior to truncation, the first 20 seconds of active speech were removed from all utterances to avoid capturing similar data across multiple utterances. For NIST 2008, the performance was evaluated using the equal error rate~(EER) and the minimum decision cost function (DCF), calculated using $\emph{C}_{miss} = 10$, $\emph{C}_{FA} = 1$, and $\emph{P}_{target} = 0.01$~\cite{NIST2008}. Outer-domain data is defined as Switchboard I, II phase I, II, III corpora, and in-domain data is defined as NIST 2004, 2005 and 2006 SRE corpora. 

We have used 13 feature-warped MFCC with appended delta coefficients and two gender-dependent UBMs containing 512 Gaussian mixtures throughout our experiments. The UBMs were trained on out-domain data, and then used to calculate the Baum-Welch statistics before training a gender dependent total-variability subspace of dimension $R_{w} = 500$. The pooled total-variability representation was trained using out-domain data. For out-domain PLDA speaker verification system, the GPLDA parameters were trained using Switchboard I, II phase I, II, III corpora. We empirically selected the number of eigenvoices~($N_{1}$) equal to 120 as best value according to speaker verification performance over an evaluation set. 150 eigenvectors were selected for LDA estimation. S-normalisation was applied for experiments. 150 NIST telephone speakers with 10 sessions were used for NIST S-normalization data, and randomly selected utterances from Switchboard I, II phase I, II, III were pooled to form the Switchboard S-normalisation dataset~\cite{Shum2010}.
\begin{table}
\caption{\label{tab:IDV_modified_IDV} \emph{Performance comparison of IDV-compensated and modified IDV-compensated PLDA systems on NIST 2008 short2-short3 evaluation condition. The best performing systems by both EER are highlighted across each row.}}
\begin{center}
\scalebox{0.9}{
\begin{tabular}{l c c} \hline
\textbf{System} & \textbf{Without Snorm} & \textbf{With Snorm} \\ \hline
\footnotesize Out-domain PLDA & 4.86\% & 3.85\%  \\
\footnotesize IDV-compensated PLDA & 4.37\% & 3.55\% \\
\footnotesize Modified IDV-compensated PLDA & \textbf{3.79\%} & \textbf{3.29\%} \\ \hline
\end{tabular} }
\end{center}
\end{table}
\section{Results and Discussions} \label{sec:results and discussions}
\subsection{Analysis of in- and out-domain PLDA speaker verification}
Previous studies have found that when speaker verification is developed in one domain and evaluated in a different domain, it significantly affects the speaker verification performance~\cite{Garcia-Romero2014}. In this section, in- and out-domain PLDA speaker verification systems are closely studied with short utterance evaluation data. 

Figure~\ref{fig:in and out domain PLDA on short eval} depicts results comparing the performance of in- and out-domain PLDA speaker verification on short utterance evaluation conditions. It can be observed from the EER performance indicated by the bars that in-domain PLDA approach shows better performance than out-domain PLDA approach in all short utterance evaluation conditions. However, it was also observed from the in-domain performance gain indicated by the red line in Figure~\ref{fig:in and out domain PLDA on short eval} that when full-length utterance is used for evaluation, in-domain PLDA approach shows more than 28\% improvement in EER and DCF values over out-domain PLDA approach and when evaluation utterance length reduces the performance gain of in-domain PLDA approach reduces in increasing rate. Further, when very short utterance are used for evaluation, there is no performance difference between in- and out-domain PLDA speaker verification as short utterance evaluation data i-vectors have more variations due to phonetic variations when compared to the dataset mismatch between in- and out-domain data. Results also suggest that data mismatch between in- and out-domain data is not influencing the performance of short utterance-based PLDA speaker verification system.
\begin{figure*}
\begin{center}
\begin{tabularx}{\linewidth}{XXX}
\subfigure[SWB data for S normalization]{
\includegraphics[width=8cm, height=5cm]{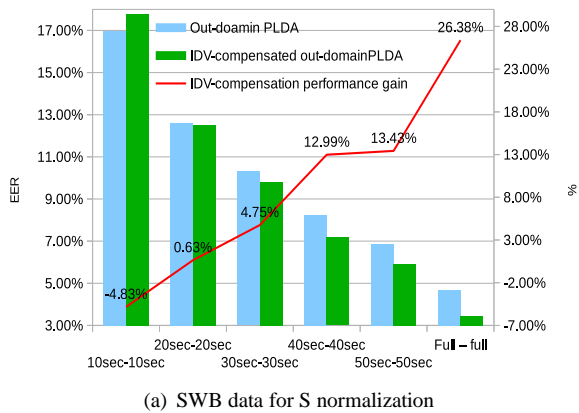}%
\label{fig:IDCN short eval snorm swb}%
} &
\subfigure[NIST data for S normalization]{
\includegraphics[width=8cm, height=5cm]{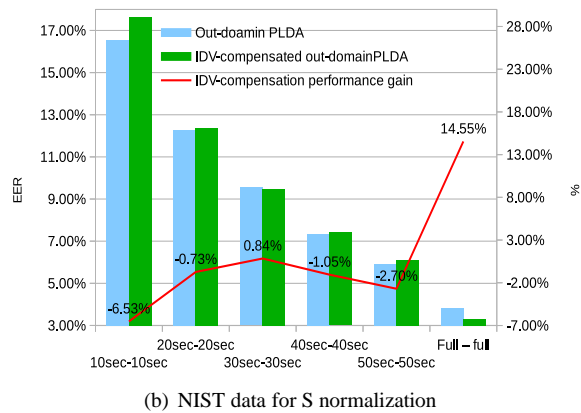}%
\label{fig:IDCN short eval snorm nist}%
}
\end{tabularx}
\end{center}
\caption{\emph{Comparison of the performance of modified IDV-compensated out-domain PLDA speaker verification against out-domain PLDA speaker verification on common set of NIST 2008 different short utterance evaluation conditions is shown in bar chart. The performance gain of modified IDV-compensated PLDA over out-domain PLDA is also shown in line graph} \subref{fig:IDCN short eval snorm swb} \emph{SWB data for S normalization,} \subref{fig:IDCN short eval snorm swb} \emph{NIST data for S normalization.}}
\label{fig:IDCN and out domain PLDA on short eval}
\end{figure*}
\begin{table}
\caption{\label{tab:score norm with full and matched length} \emph{Performance comparison of modified IDV-compensated PLDA system with full and matched length score normalization data. The best performing systems by both EER are highlighted across each row.}}
\begin{center}
\scalebox{0.9}{
\begin{tabular}{c c c} \hline
\textbf{Evaluation} &  \multicolumn{2}{c}{\textbf{S-norm development data}} \\
\textbf{utterance} &  \textbf{Full-length} & \textbf{Matched length} \\ \hline
10sec - 10 sec & \textbf{17.63\%} & 17.64\% \\
20sec - 20 sec & \textbf{12.36\%} & \textbf{12.36\%} \\
30sec - 30 sec & \textbf{9.47\%} & \textbf{9.47\%} \\
40sec - 40 sec & 7.41\% & \textbf{7.09\%} \\
50sec - 50 sec & 6.09\% & \textbf{5.85\%} \\ \hline
\end{tabular} }
\end{center}
\end{table}

\subsection{Analysis of modified IDV-compensated out-domain PLDA speaker verification}
In previous section, we have found that when speaker verification is evaluated on very short utterances, dataset mismatch doesn't influence the performance as short utterance evaluation data has a lot of uncertainties due to phonetic variation. In this section, initially both IDV and modified IDV approaches will be analysed on NIST standard conditions and subsequently the best IDV approach will be investigated with short utterance evaluation condition.

The Table~\ref{tab:IDV_modified_IDV} compares the performance of IDV-compensated and modified IDV-compensated PLDA systems against out-domain PLDA system on NIST 2008 short2-short3 evaluation conditions and it can be observed that modified IDV-compensated PLDA achieves 7\% improvement over IDV-compensated PLDA systems.

The performance comparison between modified IDV-compensated out-domain PLDA approach and out-domain PLDA approach is shown in Figure~\ref{fig:IDCN and out domain PLDA on short eval}. It has been observed with the aid of Figure~\ref{fig:IDCN and out domain PLDA on short eval}~(a) and~(b) that modified IDV-compensated PLDA shows respectively 26\% and 14\% improvement over out-domain PLDA speaker verification when SWB and NIST data are respectively used for S normalization. It was also observed with the aid of Figure~\ref{fig:IDCN and out domain PLDA on short eval}~(a) that when evaluation utterance length reduces, the performance gain by modified IDV-compensation also reduces. We believe that this is due to the fact that short utterance i-vectors have more variation arising from the phonetic content compared to variations arising from mismatch between in and out domain data which we compensate using our modified IDV.
\subsection{Analysis of PLDA speaker verification with matched length score normalization data}
Table~\ref{tab:score norm with full and matched length} presents the results comparing the performance of the modified IDV-compensated PLDA system with full-length score normalization and matched-length score normalization~(score normalization data truncated to same length as evaluation data). We found that matched-length score normalization improves the EER performance of modified IDV-compensated PLDA system when utterance length increases above 40 sec. This shows that rather than being a hindrance to normalisation performance, limited development data (if matched in length), can improve normalisation for speaker verification.
\section{Conclusion} \label{sec:conclusion}
This paper explored how the in- and out-domain PLDA speaker verification behave when enrolment and verification lengths are reduced. Our experiment studies have found that when full-length utterance were used for evaluation, in-domain PLDA approach showd more than 28\% improvement in EER and DCF values over out-domain PLDA approach and when short utterances were used for evaluation, the performance gain of in-domain speaker verification reduced at an increasing rate. The IDV and modified IDV compensations were used to compensate the mismatch between in- and out-domain data, and found that modified IDV is better approach than IDV approach. Modified IDV-compensated out-domain PLDA showed respectively 26\% and 14\% improvement over out-domain PLDA speaker verification when SWB and NIST data were respectively used for S normalization. Further when evaluation utterance length reduced, the performance gain by IDV also reduced as short utterance evaluation data i-vectors have more variations due to phonetic variations compared dataset mismatch between in- and out-domain data.
\section{Acknowledgements}
This project was supported by an Australian Research Council (ARC) Linkage grant LP130100110.

\bibliographystyle{ieeetr}
\bibliography{research}
	
\end{document}